# The Variable Optical Polarization of J0948+0022

Miller, H., Eggen, J., Maune, J.
*Georgia State University, Atlanta, GA 30303, USA*

The results of monitoring the optical polarization and the gamma-ray flux for the radio-loud narrow-line Seyfert 1 galaxy is reported. We have detected a weak correlation between the gamma-ray variability and the optical variability. However, these observations do not provide conclusive evidence that the emissions are due to a relativistic jet oriented close to the line-of-sight to the observed.

## 1. INTRODUCTION

Prior to the launch of Fermi, blazars and broad-line radio galaxies (BLRG) were the only classes of AGN known to be gamma-ray emitters. However during the first year of operation, Fermi detected gamma-ray emission from the narrow-line Seyfert 1 galaxy, PMN J0948+0022 (Abdo et al, 2009). Narrow Line Seyfert 1 galaxies (NLS1s) are a class of AGN first identified by Osterbrock and Pogge (1985). They are characterized by optical spectra which contain Balmer emission lines that are narrow with FWHM (H-beta) < 2000km/sec, strong Fe II emission, weak [OIII] 5007 emission, and generally [O III]/H-beta < 3. Although Seyfert galaxies are known to be strong and variable x-ray sources, NLS1s exhibit extraordinarily rapid, large amplitude x-ray variability (e.g., Boller et al. 1996). McHardy et al. (2006) and collaborators have shown, based on power spectral density (PSD) studies of the x-ray variability of NLS1s, that NLS1s are likely to be lower mass/higher accreting systems than normal Seyfert 1 galaxies. This is consistent with the earlier investigation of Boroson and Green (1992) and their physical interpretation of the eigenvector 1. While NLS1s are generally radio-quiet, a small subset of NLS1s (~7%) are radio loud (R>10). When one compares this with normal Seyfert 1 galaxies, ~15% of normal Seyfert 1 galaxies (Komossa et el. 2006) or AGN (Padovanni and Urry,1995) are radio loud. An even smaller percentage (~2.5%) of NLS1s are very radio loud (R>100) . PMN J0948+0022 is one of the NLS1s which is very radio-loud.

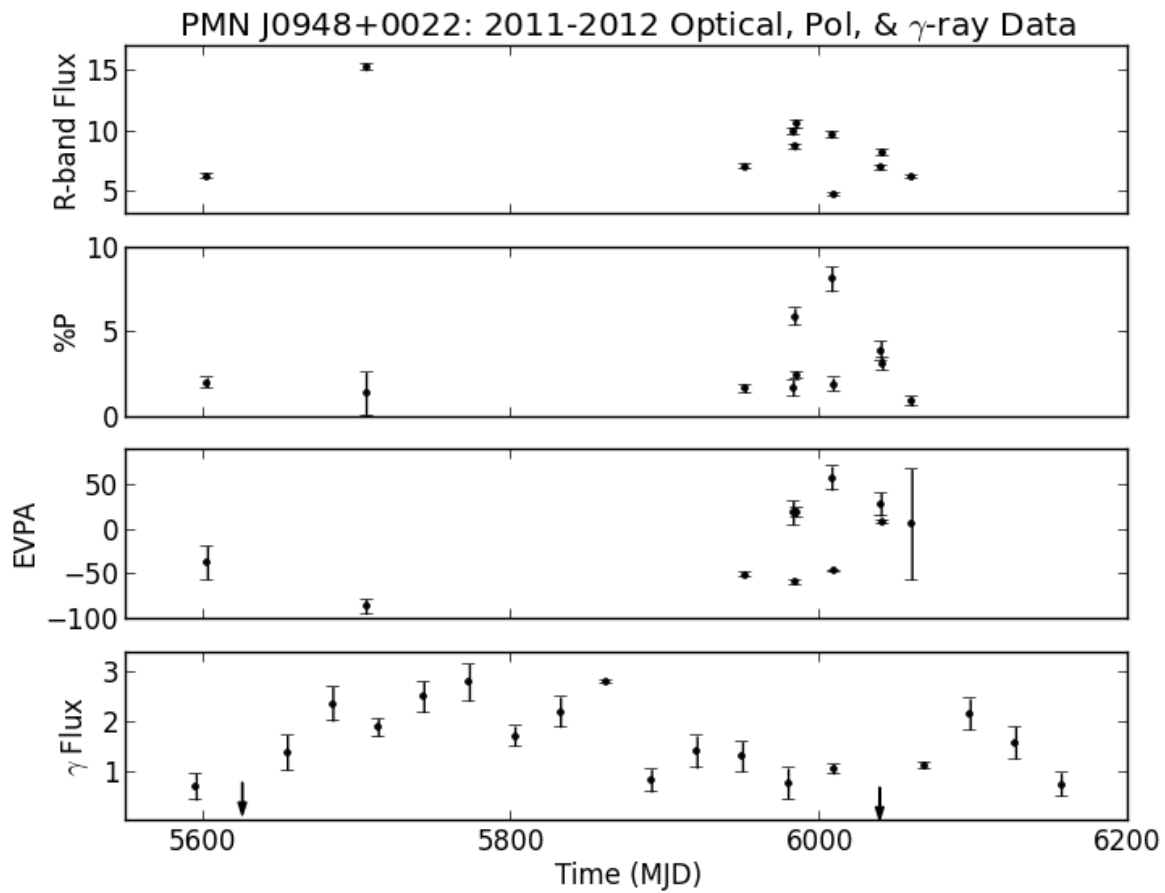

Figure 1: Recent observations of the variation of the polarization, EVPA, and γ-ray flux for PMN J0948+0022

## 2. OBSERVATIONS

### 2.1. Optical Photopolarimetry

All optical data used in this study were obtained with the 72-inch Perkins telescope at Lowell Observatory in Flagstaff, Arizona, using the PRISM instrument which includes a polarimeter with a rotating half-wave plate. Data were obtained during seven observing runs between February, 2011 and May, 2012. The specific dates of each observation are given in Table 1. The observations consisted of a series of 3-4 measurements for the Q and U Stokes parameters per object. Each series consisted of four images, each taken at different instrumental position angles – 0°, 45°, 90°, and 135° – of the waveplate. Corrections to polarimetric values were obtained from two sources: in-field comparison stars and seperately-observed polarimetric standards, both polarized and un-polarized (Schmidt et al, 1992). As the camera has a wide field of view (approx. 14' x 14'), we are able to use field stars for interstellar polarization corrections by subtracting the average percent polarization of the brightest field stars. Polarized and unpolarized standard stars are used to calibrate corrections for polarization Position Angle (P.A.) and instrumental polarization (typically less than 1%) [Jorstad et al.], respectively. The data were reduced and analyzed using in-house scripts, which utilize standard packages in the PyRAF 2.0 suite of reduction tools[footnote: PyRAF is a product of the Space Telescope Science Institute, which is operated by AURA for NASA.]. Bias frames were taken at the beginning of every night and combined into a master bias that was subtracted from each image. Flat frames were taken at least once per run, using a featureless screen inside the dome. Each position of the waveplate required its own set of flats, which would later be combined into one master flat per position angle for application to the appropriate science image. Cosmic ray cleaning was performed on all science images, with the threshold and fluxratio paramteters set to 35 and 5, respectively. Aperture photometry was then performed on the calibrated science frames on an object-by-object basis. An aperture radius of 7 arcsec was used on all images both to maximize the signal-to-noise and to maintain consistency with the optical photometry being performed on this target by Maune et al. (2013). Use of the in-field comparison stars compiled by Maune et al. allowed for the ability to obtain simultaneous measures of the absolute R-band magnitude, as well as the percent polarization (%P) and electric vector position angle (EVPA).

### 2.2. FERMI-LAT Data

Gamma-ray data were obtained through the FERMI-LAT public data server. The Large Area Telescope (LAT), on board the Fermi Gamma-ray Space Telescope, is a pair-conversion detector sensitive to gamma-rays in the 20 MeV to several hundred GeV energy range [Abdo et al.]. The instrument has worked almost continuously in all-sky-survey mode since its launch in 2009, which allows coverage of the entire gamma-ray sky approximately every 3 hours. The data were reduced and analyzed using ScienceTools v9r27p1 and instrument response functions P7SOURCE_V6. We utilized the likelihood analysis procedure as described at the FSSC website. Photon fluxes in Table 2 are calculated using data from MJD 55562 to MJD 56179 (Jan. 01, 2011 to Sep. 08, 2012).

Our data were downloaded from the FERMI website on September 8, 2012 and cover a region of interest (ROI) on the sky 20° in radius, centered on the location PMN J0948+0022 (2FGL0948.8+0020 from the Fermi 2-Year Point Source Catalog). Our gamma-ray light curve consists of 20 equally-sized bins, each of which is 29.530589 days in length so as to be centered on the new moon, which was the approximate time that our observing runs at Lowell Observatory took place. The first bin began on February 03, 2011, while the last bin ended on September 8, 2012. Only data corresponding to the Source class (evclass=2) was utilized, with a 52° cut-off rock-angle of the spacecraft. An additional cut utilizing an angle of 100° from the zenith was imposed so as to minimize the contamination due to gamma-rays coming from Earth's upper atmosphere. Photon fluxes and spectral fits were derived using an unbinned maximum likelihood analysis which was accomplished using the ScienceTool gtlike.

In order to accurately measure the flux and spectral parameters of the source, gamma-rays emitted from the background needed to accounted for. To this end, two models were used: and isotropic background model accounting for extragalactic diffuse emission and residual charged particle background, and a Galactic diffuse emission model to account for diffuse sources from within our own galaxy. The isotropic model we used was the contained in the file iso_p7v6source.txt[footnote:http://fermi.gsfc.nasa.gov/ssc/data/access/lat/Background/Models.html], while the Galactic component was given by the file gal_2year7v6_v0.fits. The normalizations of both components were left to vary freely during likelihood analysis.

In order to determine the significance of the gamma-ray signal from PMN J0948+0022, we used the Test Statistic (TS). The Test Statistic is defined as TS = $2\Delta\log$(likelihood), where likelihood refers to the likelihood ratio test as described in Mattox et al. (2006). Determining the likelihood of a given photon flux being produced by a source with a given spectral model was accomplished using the gtlike Science Tool. Our source model consisted of all the known gamma-ray point sources located within a 20° radius of 2FGLJ0948.8+0020. Initial values for all spectral parameters for these sources were taken from the LAT 2-year Point Source Catalog. Along with PMN J0948+0022 and the aforementioned background models, several point sources were allowed to vary (i.e. photon indices and normalization factors were left as free parameters) during the likelihood analysis, so as to account for the inherent variability of many gamma ray sources. The type of spectral model used for a given source was the same model used for that source in the LAT 2-year catalog. Thus, a Log Parabola model was used to describe the gamma-ray spectrum of PMN J0948+0022 in this study. Sources outside our 20° Radius of Interest (RoI) but within 25° of the target were also included in the source model, as the point spread functions of these objects could result in extra photons seeping into the RoI of our target. All parameters for these sources were fixed to their 2FGL catalog values during the analysis.

## 3. SUMMARY AND CONCLUSIONS

The optical flux and polarization observations are displayed in Fig.1 along with the gamma-ray observations. PMN J0948 exhibits significant variability in optical flux, polarization and EVPA with the maximum observed polarization reaching 8%. Although the temporal coverage is somewhat sparse, there appears to be no strong correlation between the optical and gamma-ray flux or the optical polarization. The optical and gamma-ray flux both rise between 5600 – 5800 (MJD), but the optical coverage is too sparse to allow one to claim there is a significant relationship between the variations observed in these two bands. Thus, since no major flares were observed during this observing period, these observations do not preclude the existence of a relationship between these parameters.